\documentclass[12pt]{article}
\usepackage{amsmath,epsfig,graphicx,amssymb}

\newcommand{\beq}{\begin{equation}}
\newcommand{\eeq}{\end{equation}}
\newcommand{\ben}{\begin{eqnarray}}
\newcommand{\een}{\end{eqnarray}}
\newcommand{\bes}{\begin{subequations}}
\newcommand{\ees}{\end{subequations}}
\newcommand{\bFig}{\begin{figure}}
\newcommand{\eFig}{\end{figure}}

\begin{document}

\title{Classical and Quantum Theory with A New Symmetry}
\author{Partha Ghose\footnote{partha.ghose@gmail.com} \\
Centre for Natural Sciences and Philosophy, \\ 1/AF Bidhan Nagar,
Kolkata, 700 064, India\\
and\\Centre for Philosophy and Foundations of Science,\\Darshan Sadan, E-36 Panchshila Park,
 New Delhi 110017, India}
\maketitle
\begin{abstract}
A formal symmetry between generalized coordinates and momenta is
postulated to formulate classical and quantum theories of a
particle coupled to an Abelian gauge field. It is shown
that the symmetry (a) requires the field to have dynamic degrees of
freedom and to be a connection in a non-flat space-time manifold, and (b) leads to a quantum theory free of the measurement problem. It is speculated that gravitomagnetism could be a possible source of the gauge field.

\end{abstract}

\vspace{0.2in}

\vspace{0.2in}

\section{Introduction}

Quantum theory is usually formulated by using either the canonical method or Feynman's path-integral method. The latter has the advantage of preserving all relevant symmetries whereas the former is usually regarded as breaking Poincare invariance through an explicit choice of a `time' coordinate. However, as shown by Witten and Zuckerman \cite{witten, witten2}, the canonical formalism can also be developed in a way that preserves all its relevant symmetries including Poincare invariance. This approach will be adopted to formulate a theory of classical and quantum systems with a new postulated symmetry between coordinates and momenta.

In Hamiltonian mechanics the generalized coordinates $q_i$ and $p_i$ are accorded equal status and their Lagrange and Poisson brackets are canonical invariants. This symmetry between the coordinates and momenta is lost when the particles are coupled to fields like the electromagnetic potential $A_\mu$ because the canonical four-momentum $p_\mu$ of a charged particle is then replaced by the `kinetic momentum' $\pi_\mu = p_\mu - (e/c) A_\mu$ whereas its canonical coordinates $q_\mu$ remain unchanged \cite{note}. The aim of this paper is to see under what conditions the symmetry between the coordinates and momenta can be restored in this case, and to explore its consequences in quantum mechanics.

The starting point is the classical phase space defined as the space of solutions of the classical equations. One can always, if one wishes, choose a coordinate system with a time coordinate and identify the classical solutions with the initial data in that coordinate system, but there is no necessity to make such a non-covariant choice. The notion of a `symplectic structure on phase space' is a more intrinsic concept than the idea of choosing coordinates $q_i$ and $p_i$ \cite{witten2}.

\section{Differential Geometric Preliminaries}

Differential geometry has played a very useful role in the formulation of physical theories. Since certain basic concepts of differential geometry underlie the formulation of the present paper, it will be helpful to start by recapitulating them and developing them with a view to apply them to the problem at hand. Consider the configuration space of a classical system which is generally a manifold ${\cal{M}}$ with local charts $(U, x),\, x(m, m \in U)
= q = (q_1, q_2, \cdots, q_n) \in {\bf R}^n$. One can define the tangent
vectors $X_i^q = \partial/\partial q^i, \,q \in {\bf R}^n$ and via the inverse mapping $x^{- 1}$ the tangent vectors
$X_i^m = \partial/\partial x^i, m \in U$. These tangent vectors
span the tangent space at the point $m \in U$ and are fibres
on ${\cal{M}}$. The fibres on all points on ${\cal{M}}$ together
with ${\cal{M}}$ constitute the tangent bundle $T{\cal{M}}$. The
dual to the tangent bundle is called the cotangent bundle
$T^*{\cal{M}}$ with $\pi: T^*{\cal{M}}\rightarrow {\cal{M}}$ the projection.
One can define a canonical one-form $\theta$ on $T^*{\cal{M}}$ by \cite{abraham}
\beq
\theta (\alpha)w = \alpha. T \pi (w)
\eeq
where $\alpha \in T^*{\cal{M}}$ and $w \in T_\alpha (T^*{\cal{M}})$.
The canonical two-form is defined by $\omega = - d \theta,\, d \omega = 0$. This is a reflection of the fact that $T^*{\cal{M}}$ is a symplectic manifold. If ${\cal{M}}$ is finite dimensional, the formula for $\theta$ in a local chart $(U, x)$ may be written as
$\theta = \sum_i p_i dq^i$ where the exterior derivatives $dq^i$ span the
cotangent space and are dual to $X_i^q$: $\langle dq^i, X_j^q\rangle =
\delta^i_ j$. The $p_i$ are the momenta conjugate to the coordinates
$q_i$. The two-form $\omega (q,p) = - d
\theta = \sum_i d q^i \wedge d p_i$ and it is
closed, i.e., $d \omega = 0$. It is well-known that one can always
associate a Poisson manifold $(T^*{\cal{M}},\{,\})$ with the
sympletic manifold $T^*{\cal{M}}$. The fundamental Poisson brackets
of $q_i$ and $p_j$ in a chart $(U, x)$ are \beq \{q_i,p_j\} = \delta_{i j}.\eeq
$T^*{\cal{M}}$ can be regarded as a $2n$ dimensional manifold called
`phase space' with coordinates $(q_1, \cdots, q_n, p_1,
\cdots, p_n) \in U$ rather than a bundle.

Similarly, for infinite dimensional systems like fields one
considers the manifold ${\cal{B}}$ of potentials $B_ \mu$. The
corresponding phase space is then the cotangent bundle
$T^*{\cal{B}}$ with the canonical symplectic structure. Since the
Lagrangian can be written as \beq {\cal{L}}= - \frac{1}{4
\chi}F_{B\mu \nu}F_B^{\mu \nu}, \eeq the canonical momentum is
$\pi_B^ \mu =(1/c\, \chi) F_B^{\mu 0} d^3 x = (1/c\, \chi)
F_B^{\mu \lambda}\eta^\lambda d^3 x$ with
$\eta^\lambda\eta_\lambda = -1$. The canonical symplectic
structure $\omega$ on $T^*{\cal{B}}$ is
 \beq
  \omega\left((B_{1},\pi_{B1}), (B_{2},\pi_{B2})\right) = \int_{{\bf R}^3}(\pi_{B2}.B_{1} - \pi_{B1}.B_{2}) d^3 x,
 \eeq
and the associated fundamental Poisson bracket is \beq \{F,H
\}_{(B,\, \pi_B)} = \int_{{\bf R}^3} \left(\frac{\delta F}{\delta
B}.\frac{\delta H}{\delta \pi_B} - \frac{\delta F}{\delta
\pi_B}.\frac{\delta H }{\delta B} \right) d^3 x \eeq where $\delta
F/\delta B$ is the vector field defined by \beq D_{B} F(B, \pi_B)\,
.\, B^\prime = \int \frac{\delta F}{\delta B}\, . \, B^\prime  d^3
x\eeq with the vector field $\delta F/\delta \pi_B$ defined
similarly. For further details, see \cite{abraham}.

\begin{figure}
\begin{picture}(100,100)
\put(95,45){$T_m M$}
\put(120,50){\vector(1,0){100}}
\put(165,40){$d \varphi_m$}
\put(225,45){$T_{m^\prime} M$}
\put(115,115){\vector(0,-2){60}}
\put(230,85){$\tau_{m^\prime}$}
\put(225,120){$T^*_{m^\prime} M$}
\put(120,120){\vector(1,0){100}}
\put(165,125){$d \varphi_m^*$}
\put(95,120){$T^*_m M$}
\put(225,115){\vector(0,-2){60}}
\put(100,85){$\tau_m$}
\end{picture}
\caption{The commutative diagram illustrating the pull-back map $d\varphi^*_m$}
\end{figure}

If a manifold $M$ is `curved', the tangent spaces $T_m M$ and $T_{m^\prime}M$ at two infinitesimally separated neighbouring points $m, m^\prime \in M$ are disjoint. The connection is a mapping of these tangent spaces. Let $\varphi: m \rightarrow m^\prime$ be a map. Then $d \varphi_m : T_m M \rightarrow T_{\varphi (m)} M$, i.e., the tangent vectors to $M$ at $m$ are mapped to the tangent vectors to $M$ at $m^\prime$ by the differential or covariant derivative $d \varphi_m$ which consists of the ordinary partial derivative plus the connection. This is the connection map $d \varphi_m$ (Fig. 1). Let $\tau_m: T^*_m M \rightarrow T_m M$ and
$\tau_{m^\prime}: T^*_{m^\prime} M \rightarrow T_{m^\prime} M$. Then $d \varphi_m^*: T^*_ m M \rightarrow T^*_{m^\prime} M$ is the pullback map. In terms of local charts on the manifold let the map
\beq
p \rightarrow \pi = p - \frac{2 m}{c} B \label{mom}\eeq
correspond to the pullback map $d \varphi_m^*$, and let the coordinates of the $T^*_m M$ bundle be $(\cal{Q}, \pi)$. Since $(q,p)$ and $(B, \pi_B)$ are canonical pairs, we have
\beq
{\cal{Q}} = q - \frac{c}{2 m} \pi_B.\label{coord}\eeq

One can regard $(q, {\cal{Q}})
\in V \times V$ where $V$ is a vector space and $(p\,
\otimes \pi) \in V \times V$. Then one has a nondegenerate
symplectic two-form on $V \times V \times V \times V$ given by
\ben \omega \left((q, {\cal{Q}})_1,
(p \otimes \pi)_1, (q, {\cal{Q}})_2, (p \otimes
\pi)_2\right) &=& (p_1 \otimes \pi_1)
(q_1, {\cal{Q}}_1) - (p_2 \otimes \pi_2) (q_2, {\cal{Q}}_2)\nonumber\\
&=& \pi_1 (q_1)\,.\, p_ 1 ({\cal{Q}}_1) - \pi_2 (q_2)\,.\, p_
2 ({\cal{Q}}_2). \een
Consider the phase spaces $({\cal{Q}}, p) \in V_{{\cal{Q}}} \times V_{{\cal{Q}}}^*$ and $(q, \pi) \in V_q \times V_q^*$. The total space is $V_T = V_{{\cal{Q}}} \times V_{{\cal{Q}}}^* \times V_q \times V_q^*$. Now consider the projections $P_1 V_T = V_{{\cal{Q}}} \times V_{{\cal{Q}}}^* \times V_q^* \equiv V^\prime$ and $P_2 V^\prime = V_{{\cal{Q}}} \times V_q^*$. Then $P_2 P_1 V_T  = V_{{\cal{Q}}} \times V_q^*$, and $({\cal{Q}}, \pi) \in V_{{\cal{Q}}} \times V_q^*$. This shows that the space ${\cal{Q}}$-$\pi$ is a projection of the higher dimensional phase space $V_T$ that allows Hamiltonian flows. Similarly, consider the projections $P_3 V_T = V^*_{{\cal{Q}}} \times V_{q} \times V_q^* \equiv V^{\prime \prime}$ and $P_4 V^{\prime \prime} = V_{{\cal{Q}}}^* \times V_q$. Then $P_4 P_3 V_T  = V_{{\cal{Q}}}^* \times V_q$, and $(p, q) \in V_{{\cal{Q}}}^* \times V_q$, which shows that the space $p$-$q$ is another projection of $V_T$ whose dual is the space ${\cal{Q}}$-$\pi$.

\section{Classical Theory}

With this differential geometric background let us consider the phase space of a relativistic particle coupled to a gauge field $B$. Let $Q^I, I = 1,2,...,2N$ with $Q^i = \pi^i$ for $i \leq N = 4$ and $Q^i = q^{i - N}$ for $i > N$. Then the closed two-form is $\omega (q, \pi) = \sum_i d q^i \wedge d \pi_i$ and the $2N\times 2N$ antisymmetric matrix $\omega_{IJ}$ whose non-zero matrix elements are $\omega_{i,\, i+N} = - \omega_{i+N,\,i} = 1$ is invertible. One can define the Poisson bracket of any two functions $F(Q^I)$ and $G(Q^I)$ by

\beq
[F, G] = \omega^{IJ} \frac{\partial F}{\partial Q^I}\frac{\partial G}{\partial Q^J}\eeq
where $\omega^{IJ}$ is the inverted matrix \cite{witten2}. Corresponding definitions can be given for the field $B$ following the previous section.

Let the `kinetic' momentum of the particle be

\beq \pi^\mu = p^\mu - \frac{2 m}{c}B_\mu\label{canmom}
\eeq where $2m$ is the ``charge'' of the particle.
Following the arguments of the previous section (Eqns. (\ref{mom}) and (\ref{coord})), let us define the new variable ${\cal Q}$, the `kinetic coordinate', by

\beq {\cal{Q}}^\mu = q^\mu - \frac{c}{2m}\pi_B^\mu.
\label{cancom2}\eeq Using the canonical Poisson brackets
 \ben
\{q^{\mu}, p^{\nu} \} = g^{\mu \nu},\label{pbq1}\\
\{q^{\mu}, \pi^{\nu}\} = g^{\mu \nu},\label{pbq2}\\
\{B^{\mu}, \pi_B^{\nu} \} = g^{\mu \nu},\label{pbq3}
 \een
one obtains
 \beq
 \{{\cal{Q}}^\mu, {\cal{Q}}^\nu\} = 0,
 \label{pb1}\eeq
 \beq
 \{{\cal{Q}}^\mu,\pi^\nu\} = \{q^\mu, p^\nu\} + \{\pi_B^{\mu}, B^ \nu\}= g^{\mu \nu} - g^{\mu \nu} = 0
 \label{A1}\eeq
 and
 \ben
 \{{\cal{Q}}^\mu, p^\nu \} &=& \{ q^\mu, p^\nu\} - \{\frac{c}{2 m}\pi_B^\mu, p^\nu  \}= g^{\mu \nu} .\label{pb3}
 \een It follows from these results that in addition to $(q, \pi)$ one can also choose $({\cal{Q}}, p)$ as a canonical pair.

As explained in the previous section, Eqn. (\ref{canmom}) is the momentum space representation (or pullback map) of the covariant derivative $d\phi_m$ which takes the coordinate form
\beq
D^\mu = \partial^\mu - \frac{i}{2 m \chi}B^\mu.\label{covderivative}
\eeq
with $B$ as the connection. $\pi_B$ connects the two coordinates $q$ and ${\cal{Q}}$ (Eqn. \ref{cancom2}) just as $B$ connects the two momenta $p$ and $\pi$ (Eqn. \ref{canmom}).
Just as $\pi$ is the kinematic momentum of the particle conjugate to its canonical position $q$, the Poisson bracket (\ref{pb3}) implies that ${\cal{Q}}$
is the kinematic position conjugate to the canonical momentum $p$. ${\cal{Q}}$ and $\pi$ carry global information about the configuration manifold of the particle through the connection $B$ and its conjugate $\pi_B$.

An important and well-known property of covariant derivatives is
that they do not commute, \beq [D^\mu, D^\nu] = -
\frac{i}{2m\chi}F_B^{\mu\nu},\label{hol} \eeq the commutator being
the curvature $F_B^{\mu\nu}$ of the connection. This
non-commutativity of the covariant derivatives is a {\em
classical} (i.e., non-quantum theoretic) result following from the
geometrical fact that parallel transporting a vector around a
closed loop on a curved manifold results in a different vector.
This failure to return to the initial vector is known as holonomy.
Thus, the covariant derivatives carry global information about the
manifold.

Thus, the postulated symmetry between the `kinematic' momenta and
coordinates of the particle requires the connection $B$ to be
dynamical (Eqn. \ref{pbq3}) and the space-time manifold to be
non-flat (Eqn. \ref{hol}). However, the local charts on which the equations are written are all flat.

The total Hamiltonian of the interacting system is therefore
\beq
H = H_P (p - (2m/c)A_g = \pi) + H_{GEM} (A_g, \pi_g - (2m/c)q = - (2m/c){\cal{Q}}).
\eeq
The equations of motion for the particle in terms of the variables $({\cal{Q}}, p)$ are then \beq
\dot{{\cal{Q}}}_\mu = \{{\cal{Q}}_\mu , H \}_{({\cal{Q}},p)} = \frac{\partial
H}{\partial p^\mu}, \,\,\,\,\,\,\dot{p}_\mu
= \{p_ \mu , H\}_{({\cal{Q}},p)} = - \frac{\partial H}{\partial
{\cal{Q}}^\mu},\label{HE1}\eeq
and in terms of the variables $(q, \pi)$, they are 
\beq \dot{q_\mu} = \{ q_\mu , H \}_{(q,\pi)} =
\frac{\partial H}{\partial \pi^\mu}, \,\,\,\,\,\,\dot{\pi}_\mu =
\{\pi_\mu , H\}_{(q,\pi)} = - \frac{\partial H}{\partial
q^\mu}.\label{HE2}\eeq These equations show that $(q, p)$ act as
the fundamental canonical variables ensuring Hamiltonian flows underlying the
evolution of the kinematic variables $({\cal{Q}},\pi)$. The ${\cal{Q}}$-$\pi$ space
of the particle is a projection of a higher dimensional phase space $V_T$---it is dual to the canonical phase space $p$-$q$ (see ptevious section).

\section{Quantum Theory}

It is now straightforward to construct the quantum theory of the
system by adopting the standard canonical procedure of replacing
the classical Poisson brackets (\ref{pbq1}) through (\ref{pb3}) by
commutators. One gets \beq [q^\mu, \hat{\pi}^\nu] = [q^\mu,
\hat{p}^\nu] = i\hbar\, g^{\mu \nu},\label{comm1}\eeq \beq [B^
{\mu}, \hat{\pi}_B^\nu] = i \hbar\, g^{\mu\nu},\label{comm2}\eeq
\beq [\hat{{\cal{Q}}}^\mu,\hat{\pi}^\nu] = [q^\mu,\hat{p}^\nu] +
[\hat{\pi}_B^\mu , B^\nu] = 0,\label{A2}\eeq \beq [\hat{{\cal{Q}}}^
\mu, \hat{p}^\nu] = i \hbar\, g^{\mu\nu}.\label{comm3}\eeq It
follows from these commutators that $\hat{p}^\mu (= -
i\hbar\partial^\mu$), $\hat{\pi}_B^ \mu (= - i \hbar \delta/\delta
B_\mu$), $q^\mu$, $\hat{{\cal{Q}}} \mu$ and $\hat{\pi}^\mu$ are
all hermitian. Hence, $\hat{\pi}^\mu = - i\hbar\partial^\mu -
(2m/c) B^\mu$ and therefore \beq [\hat{\pi}^\mu, \hat{\pi}^\nu] =
\frac{2 i m \hbar}{c} F_B^{\mu \nu} \label{picomm} \eeq but \beq
[\hat{{\cal{Q}}}^ \mu, \hat{{\cal{Q}}}^\nu] = 0.\label{Qcomm} \eeq
A comparison of (\ref{hol}) and (\ref{picomm}) shows that the
latter is a consequence of the curvature and holonomy of the
connection $B$. Thus, this commutator carries global information
about the configuration manifold. For example, it vanishes in flat
space-time regions where $F_B^{\mu\nu} = 0$ but $B^\mu \neq 0$ and
$\hbar \neq 0$.

As in the classical theory, $\hat{{\cal{Q}}}$ and $\hat{\pi}$ are
the `kinematic' coordinate and momentum operators respectively of
the interacting particle which carry global information about the
manifold, whereas $\hat{p}$ and $\hat{q}$ are the local canonical
momentum and coordinate operators respectively that enable
underlying Hamiltonian (or Schr\"{o}dinger) evolutions to occur.
{\em Significantly, $\hat{{\cal{Q}}}$ and $\hat{\pi}$ have
simultaneous eigenvalues because of} (\ref{A2}). This implies that
quantum theory admits `trajectories' of the particle in the
${\cal{Q}}$-$\pi$ space. One can define the density operator

\beq
\hat{\rho} = \sum_j p_j \vert \psi_j \rangle \langle \psi_j\vert, \,\,\,\,\,\,\,\,\,\,\,\,\ p_j \geq 0, \,\,\,\,\,\,\,\,\,\,\,\,\sum_j p_j = 1
\eeq
with $\vert \psi_j \rangle$ forming a complete set of states that are simultaneous eigenstates of $\hat{{\cal{Q}}}$ and $\hat{\pi}$. It satisfies the evolution equation
\beq
i \hbar \frac{\partial \hat{\rho}}{\partial t} = [H, \hat{\rho}].
\eeq
Hence, one can define the density function of the trajectories as
\beq
G({\cal{Q}}, \pi, t) = \langle {\cal{Q}}, \pi\vert \hat{\rho}(t)\vert {\cal{Q}}, \pi\rangle
\label{denfn}\eeq with \beq
\hat{\rho}(t) = U (t) \hat{\rho}(0) U^\dagger (t),\,\,\,\,\,\,\,\,\,\, U (t) = e^{- i H t/\hbar}.
\eeq

It is important to point out here that despite the existence of `trajectories' in the theory, there are important differences from the Bohm theory \cite{bohm}. The Bohm theory imposes an additional condition, namely the guidance condition on standard quantum mechanics to define trajectories in configuration space (the ``hidden variables''). The trajectories in the ${\cal{Q}}$-$ \pi$ space are, on the other hand, consequences of the postulated new symmetry. Furthermore, the `guidance condition' in the Bohm theory results in trajectories in configuration space whose initial distribution must be {\em chosen} to be identical with the quantum mechanical distribution (the ``quantum equilibrium hypothesis''). The continuity equation then ensures that this identity is preserved in time. The trajectories in $\cal{Q}$-$\pi$ space, on the other hand,
are consequences of the commutation relations (\ref{A2}) which
are preserved in time, and the distribution of the trajectories is automatically determined by the theory for all times.

\begin{flushleft}{\bf \em The Measurement Problem}
\end{flushleft}

The postulated new symmetry as well as gauge covariance imply that $\hat{{\cal{Q}}}$ and $\hat{\pi}$ rather than $\hat{q}$ and $\hat{p}$ must be regarded as the `observables' for the particle. It follows from this that the trajectories of the particles in ${\cal Q}-\pi$ space can account for the occurrence of individual stochastic events in space-time. There is no need therefore for any additional hypothesis like an external observer or `collapse/reduction' of the state vector. The theory is thus free of the so called `measurement problem'  in a manner analogous to Bohm's theory \cite{note2}. 

One might suspect that the trajectories imply conflict with the Heisenberg uncertainty relations. This is not the case for the following reason. The standard commutation relations from which the uncertainty relations are believed to follow still hold for the canonical position $q$ and momentum $p$ of the particle (\ref{comm1}) and also for $q$ and $\pi$ (\ref{comm1}) and ${\cal Q}$ and $p$ (\ref{comm3}). Although there is no such restriction on ${\cal{Q}}$ and $\pi$ (\ref{A2}) and a particle can have simultaneous sharp values of these variables, actual measurements of ${\cal{Q}}$ and $\pi$ over an ensemble of trajectories will nevertheless exhibit the statistical `scatter relation' $\Delta {\cal{Q}} \, \Delta \pi \geq \hbar/2$ because the trajectory distribution in the ensemble is determined by the density function (\ref{denfn}) at all times \cite{popper}.

\section{Concluding Remarks}

We have seen that the postulated new symmetry between the coordinates and momenta of
particles coupled to an Abelian gauge field requires the field to be a dynamical connection in a non-flat space-time manifold. An attractive feature of this new symmetry is
the natural occurrence of {\em trajectories} in ${\cal{Q}}$-$ \pi$ space in the quantum theory, which frees the theory of the measurement problem characteristic of quantum mechanics in flat space-time. Since non-flat space-time is characteristic of gravity,
it is tempting to speculate that a possible physical
interpretation of the gauge potential $B$ is gravitoelectromagnetism
(GEM) in which the canonical momentum of a particle is given by
the relation (\ref{canmom}) with $2m$ as the ``gravitational charge'' and $\chi = G$, the Newtonian gravitational constant. As shown by several authors, GEM is a
consequence of splitting the general relativistic
space-time manifold into space and time locally \cite{mashhoon1,
braginsky, jantzenetal, mashhoon2, unni}. If this interpretation is indeed possible, it would be indicative of a deep connection between quantum theory and gravity, at least in its post-Newtonian linearized form. However, it is premature to draw any conclusion about full quantum gravity from these considerations.

This paper is a sequel to ``A Relativistic Particle and Gravitoelectromagnetism'' and is an alternative approach to it in which the role of gravitomagnetism is subsidiary to the postulated new symmetry.


\begin{thebibliography}{0}
\bibitem{note}
These definitions of `canonical' and `kinetic' momenta are not universally used.
\bibitem{note2}
The trajectories are obtained as solutions of the guidance condition, which is a first-order differential equation, and hence the initial positions must be chosen such that the position distribution fits the quantum mechanical distribution. This is known as the `quantum equilibrium hypothesis'. It may not always be possible to do this for multi-particle entangled states \cite{ghose2}. There is no need of any special initial condition in the present theory because the particle distribution is determined by the density function (\ref{denfn}) at all times.
\end{thebibliography}
\end{document}